\documentclass[pra,showpacs,amsmath,amssymb,preprint,byrevtex]{revtex4}
\usepackage{bm}
\usepackage{graphicx}
\usepackage{dcolumn}

\begin{document}
\newcommand{\tr}{\mathop{\mathrm{tr}}\nolimits}
\newcommand{\adj}{\mathop{\mathrm{adj}}}
\newcommand{\diag}{\mathop{\mathrm{diag}}}
\renewcommand{\Re}{\mathop{\mathrm{Re}}}
\renewcommand{\Im}{\mathop{\mathrm{Im}}}
\newtheorem{lemma}{Lemma}
\newtheorem{proposition}{Proposition}
\newtheorem{theorem}{Theorem}
\newtheorem{corollary}{Corollary}
\newtheorem{proof}{Proof}

\preprint{}
\title{Kraus representation in the presence of initial correlations}
\author{Hiroyuki Hayashi}
\email{hayashi@hep.phys.waseda.ac.jp}
\author{Gen Kimura}
\email{gen@hep.phys.waseda.ac.jp}
\author{Yukihiro Ota}
\email{ota@suou.waseda.jp}
\affiliation{Department of Physics, Waseda University, Tokyo 169--8555, Japan}
\date{\today}
\begin{abstract}
We examine the validity of the Kraus representation in the presence of initial correlations and show that it is assured only when a joint dynamics is locally unitary.
\end{abstract}
\maketitle

\section{Introduction}

In open quantum systems where time evolutions are not necessarily unitary, the following form of a map is usually considered to give the most general dynamical map (quantum operation):
\begin{equation}\label{eq:KrausRep}
\rho \to \sum_{\mu} W_{\mu} \rho W^\dagger_{\mu} \quad (\sum_{\mu}W_{\mu}^\dagger W_{\mu} = \mathbb{I}),
\end{equation}
which is known as the Kraus representation (Kraus Rep.) \cite{ref:Kraus,ref:Alicki}. This is based on the reduced dynamics with no initial correlations: A system of interest (system A) is regarded as a subsystem interacting with an environment (system B), and the total system (composite system of A and B) evolves unitarily. If there are no initial correlations, i.e., if the initial density operator $\rho_{\rm{AB}}$ of the total system is {\it factorable} \cite{ref:Buzek2}
\begin{equation}
\rho_{\rm{AB}} = \rho_{\rm{A}}\otimes \rho_{\rm{B}},
\end{equation}
one can show \cite{ref:Preskill} that a dynamical map for system A takes a form of the Kraus Rep. (\ref{eq:KrausRep}).

Recently, \v{S}telmachovi\v{c} and Bu\v{z}ek \cite{ref:Stelmchovic} showed that a map based on the reduced dynamics in the presence of initial correlations \cite{ref:IniCor} is not described by the form of Eq.~(\ref{eq:KrausRep}). In this case, an additional inhomogeneous part $\delta\rho$ appears:
\begin{equation}\label{eq:KrausRep+D}
\rho \to \sum_{\mu} W_{\mu} \rho W^\dagger_{\mu} + \delta\rho\quad (\sum_{\mu}W_{\mu}^\dagger W_{\mu} = \mathbb{I}).
\end{equation}
This result has motivated the reconsideration of the general form of a dynamical map based on the reduced dynamics \cite{ref:Stelmchovic,ref:CP2}.

On the other hand, in the comment on Ref.~\cite{ref:Stelmchovic}, it has been pointed out by Salgado and S\'anchez-G\'omez \cite{ref:Salgado} that whether the Kraus Rep. is valid or not depends, not only on initial correlations, but also on the variety of the joint dynamics: They showed that in the case of a local unitary evolution, the dynamical map still has a form of the Kraus Rep. even in the presence of any initial correlation.

In this paper, we investigate the conditions to use the Kraus Rep. in the presence of initial correlations in detail. In Sec.~\ref{sec:RV} we review and clarify the arguments in Refs.~\cite{ref:Stelmchovic} and \cite{ref:Salgado}, introducing the operator $\rho_{\text{{\tiny COR}}}$ which quantify a correlation between systems A and B. In the main part of the paper, Sec.~\ref{sec:KRinIC}, we prove that the joint dynamics which provides the Kraus Rep. in the presence of any initial correlation is restricted to be locally unitary in $2\times M$ composite system (Sec.~\ref{sec:KRinIC2}) and $N \times M$ composite system (Sec.~\ref{sec:KRinIC3}). Sec.~\ref{sec:Conclusion} contains some conclusions and discussions.

\section{Dynamical map in the presence of initial correlations}\label{sec:RV}
First, we review \cite{not:Clarification} the results in Refs.~\cite{ref:Stelmchovic} and \cite{ref:Salgado}. In order to clarify the arguments, we introduce the following operator:
\begin{equation}\label{eq:CorrelationOp}
\rho_{\text{\tiny{COR}}} \equiv \rho_{\rm{AB}} - \rho_{\rm{A}} \otimes \rho_{\rm{B}},
\end{equation}
where $\rho_{\rm{AB}}$ is the density operator of the total system, $\rho_{\rm{A}} \equiv \tr_{\rm{B}}\rho_{\rm{AB}}$ and $\rho_{\rm{B}} \equiv \tr_{\rm{A}} \rho_{\rm{AB}}$ (reduced density operators). Notice that if there are no correlations between systems A and B, $\rho_{\text{\tiny{COR}}} = 0$, and vice versa. Therefore we call it the correlation operator hereafter. It has the following property
\begin{equation}\label{eq:TrBCor=TrACor=0}
\tr_{\rm{B}} \rho_{\text{\tiny{COR}}} = 0.
\end{equation}

The joint dynamics is described by a unitary operator $U_{\rm{AB}}(t) = \exp (-i H_{\rm{AB}} t)$ where $H_{\rm{AB}}$ is a total Hamiltonian, which can be decomposed into
\begin{equation}\label{eq:TotalHami}
H_{\rm{AB}} = H_{\rm{A}}\otimes \mathbb{I}_{\rm{B}} + \mathbb{I}_{\rm{A}} \otimes H_{\rm{B}} + V.
\end{equation}
By making use of the initial correlation operator $\rho_{\text{\tiny{COR}}}(0) \equiv \rho_{\rm{AB}}(0) - \rho_{\rm{A}}(0) \otimes \rho_{\rm{B}}(0)$ ($\rho_{\rm{A}}(0) \equiv \tr_{\rm{B}}\{\rho_{\rm{AB}}(0)\},\  \rho_{\rm{B}}(0) \equiv \tr_{\rm{A}} \{\rho_{\rm{AB}}(0)\} $) for the initial density operator $\rho_{\rm{AB}}(0)$, and on the basis of the reduced dynamics, a dynamical map for system A is obtained \cite{ref:Stelmchovic}:
\begin{eqnarray}
\rho_{\rm{A}}(t) &=& \tr_{\rm{B}}\{ U_{\rm{AB}}(t) \rho_{\rm{AB}}(0) U_{\rm{AB}}^\dagger(t)\} \nonumber \\
&=& \tr_{\rm{B}} \{U_{\rm{AB}}(t) \rho_{\rm{A}}(0) \otimes \rho_{\rm{B}}(0) U_{\rm{AB}}^\dagger(t)\} \nonumber \\
&& {}+ \tr_{\rm{B}} \{U_{\rm{AB}}(t) \rho_{\text{\tiny{COR}}}(0) U_{\rm{AB}}^\dagger(t)\} \nonumber \\
&\equiv& \sum_{\mu,\nu}M_{\mu\nu}(t)\rho_{\rm{A}}(0) M^\dagger_{\mu\nu}(t)+ \delta \rho_{\rm{A}} (t),\label{eq:Kraus+DR}
\end{eqnarray}
where $M_{\mu\nu}(t) \equiv \langle \mu | \sqrt{p_\nu}U_{\rm{AB}}(t) | \nu \rangle$ ($\rho_{\rm{B}}(0) = \sum_{\nu}p_\nu | \nu \rangle \langle \nu |$). The first term on the rightmost-hand side of Eq.~(\ref{eq:Kraus+DR}) is the Kraus Rep. ($\sum_{\mu\nu}M^\dagger_{\mu\nu}(t)M_{\mu\nu}(t) = \tr_{\rm{B}}\{ \rho_{\rm{B}}(0)\} \mathbb{I}_{\rm{A}}= \mathbb{I}_{\rm{A}}$); in addition, an inhomogeneous part which depends on the initial correlation $\rho_{\text{\tiny{COR}}}(0)$ appears:
\begin{equation}\label{eq:InHom}
\delta{\rho}_{\rm{A}}(t) \equiv \tr_{\rm{B}} \{U_{\rm{AB}}(t) \rho_{\text{\tiny{COR}}}(0) U_{\rm{AB}}^{\dagger}(t)\}.
\end{equation}
Therefore the most general map based on the reduced dynamics takes the form of Eq.~(\ref{eq:KrausRep+D}) (Kraus Rep. + inhomogeneous part (\ref{eq:InHom})).

On the other hand, in the case of a local unitary evolution, i.e., when the operator $U_{\rm{AB}}(t)$ is factorized as
\begin{equation}
U_{\rm{AB}}(t) = U_{\rm{A}}(t)\otimes U_{\rm{B}}(t),
\end{equation}
in terms of local unitary operators $U_{\rm{A}}(t)$ and $U_{\rm{B}}(t)$, the inhomogeneous part (\ref{eq:InHom}) vanishes at any time $t\in \mathbb{R}$ and for any initial correlation \cite{ref:Salgado}. Indeed, we can prove it as
\begin{eqnarray}
\delta{\rho}_{\rm{A}}(t) &=& \tr_{\rm{B}} \{U_{\rm{A}}(t)\otimes U_{\rm{B}}(t) \rho_{\text{\tiny{COR}}}(0) U_{\rm{A}}^{\dagger}(t)\otimes U_{\rm{B}}^{\dagger}(t)\} \nonumber \\
&=& U_{\rm{A}}(t) \tr_{\rm{B}}\{ U_{\rm{B}}(t) \rho_{\text{\tiny{COR}}}(0)U_{\rm{B}}^{\dagger}(t)\} U_{\rm{A}}^{\dagger}(t) \nonumber \\
&=& U_{\rm{A}}(t) \tr_{\rm{B}}\{\rho_{\text{\tiny{COR}}}(0)\} U_{\rm{A}}^{\dagger}(t) = 0,
\end{eqnarray}
where use has been made of the cyclic property of the trace and Eq.~(\ref{eq:TrBCor=TrACor=0}). Thus one obtains the following proposition \cite{not:P1}:
\begin{proposition}{\rm \cite{ref:Salgado}}\label{thm:Drho=0<=LUO}\begin{equation*}
\mathrm{Local \ unitary \ evolution} \Rightarrow \mathrm{Kraus \ Rep.},\ \forall t \in \mathbb{R},\ \rm{and} \ \forall \rho_{\text{\rm{\tiny{COR}}}}(0).
\end{equation*}
\end{proposition}
Observing this result, Salgado and S\'anchez-G\'omez \cite{ref:Salgado} assert that not only the role of the initial correlations but also that of the joint dynamics is relevant to whether a dynamical map is given by the Kraus Rep. or not. Although they have investigated only the case of a local unitary evolution, one can see that their perspective is in some sense correct: We can show that for a given joint dynamics which is not necessarily locally unitary, a particular choice of initial correlation can bring us with the Kraus Rep. Here we only illustrate this fact by using a simple model for the controlled-NOT gate \cite{ref:Nielsen,ref:Stelmchovic}, in which both systems A and B are $2$-level systems, described by the Hamiltonian
\begin{equation}
H_{\rm{AB}}= \sigma_1 \otimes \frac{1}{2}(\mathbb{I}_{\rm{B}}-\sigma_3)+\mathbb{I}_{\rm{A}} \otimes \frac{1}{2}(\mathbb{I}_{\rm{B}}+\sigma_3),
\end{equation}
where $\sigma_i$'s are Pauli spin operators (or orthogonal generators of $SU(2)$). Needless to say, this does not yield a local unitary evolution, since the non-zero interaction Hamiltonian
\begin{equation}\label{eq:VofCNOT}
V=-\frac{1}{2}\sigma_{1} \otimes \sigma_{3}
\end{equation}
is included in $H_{\rm{AB}}$. Taking into account of the property (\ref{eq:TrBCor=TrACor=0}), we expand the initial correlation operator $\rho_{\text{\tiny{COR}}}(0)$ as \cite{not:COandCP}
\begin{equation}
\rho_{\text{\tiny{COR}}}(0) = \sum_{i,j=1}^{3}\gamma^\prime_{ij}\sigma_i\otimes \sigma_j \ (\gamma^\prime_{ij} \in \mathbb{R}).
\end{equation}
It is then easy to calculate the inhomogeneous part (\ref{eq:InHom}) for the model (11),
\begin{equation}\label{eq:InHomoCNOT}
\delta\rho_{\rm{A}}(t) = 2 ({\gamma^{\prime}}_{23} \sin^2{t}+{\gamma^{\prime}}_{33} \sin{t} \cos{t}) \sigma_2  + 2({\gamma^{\prime}}_{33} \sin^2{t}-{\gamma^{\prime}}_{23} \sin{t} \cos{t}) \sigma_3.
\end{equation}
One sees that only $\gamma^\prime_{23}$ and $\gamma^\prime_{33}$ of $\gamma^\prime_{ij}$'s appear here, and therefore, inhomogeneous part (\ref{eq:InHomoCNOT}) vanishes if $\gamma^\prime_{23}=\gamma^\prime_{33}=0$ in the initial correlation operator (13), i.e.,
\begin{equation}\label{eq:ParticularIC}
\rho_{\text{\tiny{COR}}}(0) = \sum_{i,j=1}^{3}\gamma^\prime_{ij}\sigma_i\otimes\sigma_j \quad \rm{with} \quad \gamma^\prime_{23}=\gamma^\prime_{33}=0.
\end{equation}
Thus, it is shown that the dynamical map for system A reduced from the total system (controlled-NOT gate) can be described by the Kraus Rep. if a particular initial correlation (\ref{eq:ParticularIC}) is adopted. In this manner, for a given joint dynamics, it does not seem impossible to use the Kraus Rep. even in the presence of a restricted initial correlation like Eq.~(\ref{eq:ParticularIC}).

Conversely, however, the existence of such a kind of initial correlation, as that with $\gamma^{\prime}_{23}$ or $\gamma^{\prime}_{33}$ $\neq 0$ for the case of the controlled-NOT gate, is not compatible with the Kraus Rep., while in the case of the local unitary evolution, no conditions on the initial correlations are required for the Kraus Rep. to be valid, as in Proposition \ref{thm:Drho=0<=LUO}. Our purpose in this paper is to examine whether there are other joint dynamics than the local unitary evolution, compatible with the Kraus Rep. in the presence of any initial correlation.

\section{Kraus representation in the presence of initial correlations}\label{sec:KRinIC}

\subsection{Necessary Condition to use the Kraus Representation}

We will start from the following lemma:
\begin{lemma}\label{lem:Drho=>[V,COR]=0}
For any initial correlation $\rho_{\text{\rm{\tiny{COR}}}}(0)$,
\begin{equation}
\mathrm{Kraus \ Rep.},\  \forall t \in \mathbb{R}
 \Rightarrow \tr_{\rm{B}} [V,\rho_{\text{\rm{\tiny{COR}}}}(0)] = 0.
\end{equation}
\end{lemma}

{\it Proof}

\noindent Suppose that the Kraus Rep. is valid at any time $t \in \mathbb{R}$ in the presence of $\rho_{\text{\tiny{COR}}}(0)$, i.e.,
\begin{eqnarray}\label{eq:InHom=0}
\delta{\rho}_{\rm{A}}(t) = \tr_{\rm{B}} \{e^{-iH_{\rm{AB}}t} \rho_{\text{\tiny{COR}}}(0) e^{iH_{\rm{AB}}t}\} = 0.
\end{eqnarray}
By differentiating this with respect to $t$, we obtain
\begin{equation}\label{eq:DifInHom}
\tr_{\rm{B}} [H_{\rm{AB}},\rho_{\text{\tiny{COR}}}(0)] = 0,
\end{equation}
at $t=0$. Since $\tr_{\rm{B}} [H_{\rm{A}}\otimes \mathbb{I}_{\rm{B}},\rho_{\text{\tiny{COR}}}(0)] = [H_{\rm{A}},\tr_{\rm{B}}\{ \rho_{\text{\tiny{COR}}}(0)\}] = 0$ by Eq.~(\ref{eq:TrBCor=TrACor=0}) and $\tr_{\rm{B}} [\mathbb{I}_{\rm{A}}\otimes H_{\rm{B}},\rho_{\text{\tiny{COR}}}(0)] = 0$ from the cyclic property of trace, we have\begin{equation}\label{eq:TrB[VC]=0}
\tr_{\rm{B}} [V,\rho_{\text{\tiny{COR}}}(0)] = 0.
\end{equation}
\hfill{Q.E.D.}

In the following, we restrict ourselves to discussion of the composite system with which the associated Hilbert space is of finite dimension, i.e., $N \times M$ composite system with arbitrary $N, M \ge 2$. In this case, any operator can be expanded in terms of $\mathbb{I}_{\rm{A}}\otimes \mathbb{I}_{\rm{B}}$, $\sigma_i\otimes \mathbb{I}_{\rm{B}}$'s, $\mathbb{I}_{\rm{A}}\otimes \tau_j$'s and $\sigma_i\otimes \tau_j$'s ($i=1,\ldots, N^2-1$, $j=1,\ldots,M^2-1$), where $\sigma_i$'s and $\tau_k$'s are the orthogonal generators of $SU(N)$ and $SU(M)$ \cite{ref:Mahler}, respectively, and satisfy
\begin{subequations}\label{eq:gen}
\begin{eqnarray}
\sigma_i &=& \sigma^\dagger_i, \ \tr_{\rm{A}} \sigma_i = 0, \ \tr_{\rm{A}} \{\sigma_i\sigma_j\} = 2\delta_{ij}, \\
\tau_k &=& \tau^\dagger_k, \ \tr_{\rm{B}} \tau_k = 0, \ \tr_{\rm{B}} \{\tau_k\tau_l\} =2 \delta_{kl}.
\end{eqnarray}
\end{subequations}
The interaction Hamiltonian $V$ can be expanded as \begin{equation}\label{eq:ExpV}
V = \sum_{i=1}^{N^2-1}\sum_{j=1}^{M^2-1} v_{ij} \sigma_i\otimes \tau_j \quad (v_{ij} \in \mathbb{R}),
\end{equation}
since the terms linear in $\mathbb{I}_{\rm{A}}\otimes \mathbb{I}_{\rm{B}}$, $\sigma_i\otimes \mathbb{I}_{\rm{B}}$'s, $\mathbb{I}_{\rm{A}}\otimes \tau_j$'s are included in $H_{\rm{A}}\otimes \mathbb{I}_{\rm{B}}$ and $\mathbb{I}_{\rm{A}}\otimes H_{\rm{B}}$ in $H_{\rm{AB}}$.

Let us consider the following form of the initial correlation \cite{ref:gamma'}
\begin{equation}\label{eq:IniNM}
\rho^{l,m}_{\text{\tiny{COR}}}(0) \propto \sigma_l \otimes \tau_m \ (l=1,\ldots,N^2-1, m=1,\ldots,M^2-1),
\end{equation}
and suppose that the Kraus Rep. is still valid in the presence of this initial correlation. Then, from Lemma \ref{lem:Drho=>[V,COR]=0}, it follows that
\begin{eqnarray}\label{eqn:CalCondi}
&&\tr_{\rm{B}} [V,\rho^{l,m}_{\text{\tiny{COR}}}(0)] = 0 \nonumber \\
\Leftrightarrow && \sum_{i=1}^{N^2-1}\sum_{j=1}^{M^2-1} \tr_{\rm{B}} [v_{ij} \sigma_i\otimes \tau_j,\sigma_l\otimes\tau_m] \nonumber \\
&=& \sum_{i=1}^{N^2-1}\sum_{j=1}^{M^2-1}  v_{ij} ( [\sigma_i,\sigma_l]\tr_{\rm{B}}\{\tau_j\tau_m\} + \sigma_l\sigma_i \tr_{\rm{B}}[\tau_j,\tau_m])\nonumber \\
 &=& \sum_{i,n=1}^{N^2-1} 4i  v_{im} g_{iln}\sigma_n = 0,
\end{eqnarray}
where $g_{iln}$ is a completely antisymmetric structure constant of $SU(N)$: $[\sigma_i,\sigma_l] = 2i\sum_{n=1}^{N^2-1}g_{iln} \sigma_n$. Since the $\sigma_n$'s are linearly independent, we obtain
\begin{equation}\label{eq:NecCondi}
\sum_{i=1}^{N^2-1} v_{im} g_{iln} = 0,
\end{equation}
for all $n=1,\ldots,N^2-1$.

\subsection{$2 \times M$ composite systems}\label{sec:KRinIC2}

Let us consider the case of $N=2$, i.e., a $2$-level system interacting with an arbitrary $M$-level system. Notice that the structure constant of $SU(2)$ is $g_{iln} = \epsilon_{iln}$ or $g_{iln} = - \epsilon_{iln}$ \cite{ref:Gen1} ($\epsilon_{iln}$ : the Levi-Chivita symbol). Multiplying Eq.~(\ref{eq:NecCondi}) by $\epsilon_{opn}$ and summing $n$ up from $1$ to $3$, we have
\begin{eqnarray}
\sum_{i,n=1}^{3} v_{im} \epsilon_{opn}\epsilon_{iln} = 0 \Leftrightarrow \delta_{pl}v_{om} = \delta_{ol}v_{pm},
\end{eqnarray}
for any $o,p = 1,2,3$. By putting $o=l$, it follows that
\begin{equation}\label{eq:CondiForV}
 p\neq l \Rightarrow v_{pm} = 0.
\end{equation}
This result shows that in the presence of the initial correlation $\rho^{l,m}_{\text{\tiny{COR}}}(0)$ (\ref{eq:IniNM}) the interaction Hamiltonian $V$ is restricted to the form
\begin{equation}\label{eq:RV}
V = v_{lm} \sigma_l \otimes \tau_m + \sum_{i=1}^{3}\sum_{j=1,j\neq m}^{M^2-1} v_{ij} \sigma_i\otimes \tau_j,
\end{equation}
in order that the reduced dynamics be described by the Kraus Rep. Furthermore, if the Kraus Rep. is to appear under suitable choices of initial correlations, the interaction Hamiltonian $V$ is shown to vanish. Indeed, we can choose a set of initial correlations of the form (\ref{eq:IniNM}) with $2$ different $l$'s for any $m=1,\ldots, M^2-1$ \cite{not:VofCNOT}, e.g.,
\begin{equation}\label{eqn:EGini}
\{ \ \rho^{1,1}_{\text{\tiny{COR}}}(0),\rho^{2,1}_{\text{\tiny{COR}}}(0), \rho^{1,2}_{\text{\tiny{COR}}}(0),\rho^{2,2}_{\text{\tiny{COR}}}(0), \cdots, \rho^{1,M^2-1}_{\text{\tiny{COR}}}(0),\rho^{2,M^2-1}_{\text{\tiny{COR}}}(0) \ \}
\end{equation}
as such initial correlations. It is evident from the statement (\ref{eq:CondiForV}) that in order to keep the validity of the Kraus Rep. as the reduced dynamics under any one of initial correlations (\ref{eqn:EGini}) the interaction Hamiltonian $V$ has to vanish. The case of $V=0$ corresponds to that of local unitary evolution, since the total Hamiltonian $H_{\rm{AB}}$ becomes $H_{\rm{AB}} = H_{\rm{A}}\otimes \mathbb{I}_{\rm{B}} + \mathbb{I}_{\rm{A}} \otimes H_{\rm{B}}$ and $U_{\rm{AB}}(t) = \exp (-i H_{\rm{AB}} t) = \exp (-iH_{\rm{A}} t)\otimes \exp(-iH_{\rm{B}} t) = U_{\rm{A}}(t)\otimes U_{\rm{B}}(t)$. Therefore, we obtain
\begin{proposition}\label{cor:2M}
If the dynamical map for a $2$-level system reduced from any $2 \times M$ composite system is to take the form of the Kraus Rep. in the presence of the specific initial correlations like in Eq.~(\ref{eqn:EGini}), the joint dynamics is locally unitary.
\end{proposition}
The result is not directly generalized to the case of $N \ge 3$, because the argument for $N=2$ is dependent on particular property of the Levi-Chivita symbol, while various structure constants appear for $N \ge 3$. However, as will be shown in the next section, if the condition is strengthened to cover any initial correlation, the same conclusion as in Proposition \ref{cor:2M} follows, i.e., the converse statement of Proposition \ref{thm:Drho=0<=LUO} is also satisfied.

\subsection{$N \times M$ composite systems}\label{sec:KRinIC3}

We prove \cite{not:Inf} that a joint dynamics is restricted to be locally unitary if the dynamical map for an $N$-level system reduced from an $N\times M$ composite system is still to be described by the Kraus Rep. even in the presence of any initial correlation, i.e., the converse of Proposition \ref{thm:Drho=0<=LUO} holds.
\begin{proposition}\label{thm:Drho=0=>LUO}
\begin{equation*}
\mathrm{Kraus \ Rep.},\  \forall t \in \mathbb{R},\
\rm{and} \   \forall \rho_{\text{\rm{\tiny{COR}}}}(0) \Rightarrow \mathrm{local \ unitary \ evolution}.
\end{equation*}
\end{proposition}

{\it Proof}

\noindent
From Lemma \ref{lem:Drho=>[V,COR]=0}, the interaction Hamiltonian $V$ satisfies the condition (\ref{eq:TrB[VC]=0}) for any $\rho_{\text{\tiny{COR}}}(0)$. We will show that such a $V$ has to vanish.

By using the singular value decomposition \cite{ref:Nicholson} for $v_{ij}$ in Eq.~(\ref{eq:ExpV}), the coefficient matrix $v_{ij}$ is diagonalized with a suitable choice of generators $\sigma^{\prime}_i$'s and $\tau^\prime_j$'s which satisfy Eqs.~(\ref{eq:gen}). In these bases $V$ is expanded simply as
\begin{equation}
V = \sum_{k=1}^{L^2-1}v_{k} \sigma^\prime_k\otimes \tau^\prime_k \quad (v_k \ge 0),
\end{equation}
where $L \equiv \mathrm{min}[N,M]$. Since the condition (\ref{eq:TrB[VC]=0}) holds for any $\rho_{\text{\tiny{COR}}}(0)$, we can choose $(N^2-1)\times(M^2-1)$ correlation operators $\rho^{l,m}_{\text{\tiny{COR}}}(0) \propto \sigma^\prime_l\otimes\tau^\prime_m \ (l=1,\ldots,N^2-1,\ m=1,\ldots,M^2-1)$ \cite{ref:gamma'} as $\rho_{\text{\tiny{COR}}}(0)$ in Eq.~(\ref{eq:TrB[VC]=0}). It follows that
\begin{eqnarray}
\tr_{\rm{B}} [V,\rho^{l,m}_{\text{\tiny{COR}}}(0)] &\propto& \sum_{k=1}^{L^2-1} \tr_{\rm{B}} [v_{k} \sigma^\prime_k\otimes \tau^\prime_k,\sigma^\prime_l\otimes\tau^\prime_m] \nonumber \\
&=& \sum_{k=1}^{L^2-1} v_k ( [\sigma^\prime_k,\sigma^\prime_l]\tr_{\rm{B}}\{\tau^\prime_k\tau^\prime_m\} + \sigma^\prime_l\sigma^\prime_k    \tr_{\rm{B}}[\tau^\prime_k,\tau^\prime_m])\nonumber \\
 &=& \sum_{n=1}^{N^2-1} 4i v_m g^{\prime}_{mln}\sigma^\prime_n,
\end{eqnarray}
where $g^{\prime}_{mln}$ is the structure constant:
\begin{equation}\label{eq:SC}
[\sigma^\prime_m,\sigma^\prime_l] = 2i\sum_{n=1}^{N^2-1}g^\prime_{mln} \sigma^\prime_n.
\end{equation}
From the condition (\ref{eq:TrB[VC]=0}) and linear independence of generators $\sigma^\prime_n$'s, we obtain
\begin{equation}\label{eq:lastcondition}
v_m g^\prime_{mln}= 0
\end{equation}
for all $m = 1,\ldots,L^2-1$ and $\ l,n = 1,\ldots,N^2-1$. Assume that $v_k \neq 0$ for some $k=1,\ldots,L^2-1$, then $g^{\prime}_{kln} = 0$ for any $l,n=1,\ldots, N^2-1$ by the condition (\ref{eq:lastcondition}); this leads to $[\sigma^\prime_k,\sigma^\prime_l] = 0$ for any $l=1,\ldots, N^2-1$ from Eq.~(\ref{eq:SC}), that is, $\sigma^\prime_k$ commutes with any other operator. Therefore, $\sigma^\prime_k = c \mathbb{I}_{\rm{A}} \ ( c \in \mathbb{C})$. However, this contradicts the linear independence of $\sigma^\prime_k$ and $\mathbb{I}_{\rm{A}}$. In consequence, we obtain $v_m = 0$ for any $m$, i.e., $V = 0$, which means the joint dynamics is locally unitary.

\hfill{Q.E.D.}

\section{Conclusions and discussions}\label{sec:Conclusion}

We have investigated the role of the initial correlations in the reduced dynamics, focusing on the validity of the Kraus Rep. It has been shown that the joint dynamics is restricted to be locally unitary if the Kraus Rep. is valid as the reduced dynamics in the presence of any initial correlation (Proposition \ref{thm:Drho=0=>LUO}). Combined with Proposition \ref{thm:Drho=0<=LUO}, we obtain
\begin{theorem}\label{thm:Drho=0<=>LUO}
The dynamical map for an $N$-level system reduced from an $N \times M$ composite system with an arbitrary initial correlation takes the form of the Kraus Rep. if and only if the joint dynamics is locally unitary:
\begin{equation*}
\mathrm{Kraus \ Rep.}, \ \forall t \in \mathbb{R},\ \rm{and} \ \forall \rho_{\text{\rm{\tiny{COR}}}}(0)  \Leftrightarrow  \mathrm{local \ unitary \ evolution}.
\end{equation*}
\end{theorem}

In quantum information theory, local unitary evolutions are important tools for, e.g., the quantum teleportation \cite{ref:Nielsen}. In that case, the Kraus Rep., although it is merely a unitary map \cite{not:P1}, is valid even in the presence of any initial correlation, as is shown in Ref.~\cite{ref:Salgado}. However, if there is an interaction between a composite system and if the initial correlation exists, we cannot generally use the Kraus Rep. \cite{ref:Stelmchovic}. In particular, the Kraus Rep. is shown to be compatible with the existence of any initial correlation, only when the joint dynamics is locally unitary (Proposition \ref{thm:Drho=0=>LUO}). It is true that even through a particular choice of the initial correlation such as in Eq.~(\ref{eq:ParticularIC}) allows one to use the Kraus Rep. for a joint dynamics which is not locally unitary. Such a case, however, will require a complete controllability of initial correlations in experiments. Therefore it would be more fruitful to accept the dynamical map (\ref{eq:KrausRep+D}) as it is with the inhomogeneous part (\ref{eq:InHom}) and to consider its application, rather than to persist in the form of the Kraus Rep. (or complete positivity \cite{ref:CP2}). Indeed, it has been pointed out \cite{ref:Stelmchovic} that the existence of the inhomogeneous part often considerably changes the dynamics, from which we expect some applications for exploring unknown quantum operations \cite{ref:Nielsen}.

We acknowledge useful comments and valuable advices from Prof. I. Ohba and Prof. H. Nakazato. One of the authors (G.K.) thanks Dr. J. Shimamura for the helpful discussions and fruitful comments.


\begin{thebibliography}{10}
\bibitem{not:Clarification}
Our review, although the rigorous proof is not given, does not depend on the finiteness of dimensionality of the associated Hilbert space, while the arguments in Refs.~\cite{ref:Stelmchovic} and \cite{ref:Salgado} are essentially restricted to finite-dimensional cases.
\bibitem{not:P1}
We notice that the dynamical map for system A reduced from the local unitary evolution is merely a unitary map: $\rho_{\rm{A}}(t) = \sum_{\mu,\nu}M_{\mu\nu}(t)\rho_{\rm{A}}(0) M^\dagger_{\mu\nu}(t) = \sum_{\mu,\nu} p_\nu \langle \mu | U_{\rm{B}}(t) | \nu \rangle \langle \nu | U^\dagger_{\rm{B}}(t) | \mu \rangle U_{\rm{A}}(t) \rho_{\rm{A}}(0) U^\dagger_{\rm{A}}(t) = U_{\rm{A}}(t) \rho_{\rm{A}}(0) U^\dagger_{\rm{A}}(t)$.
\bibitem{not:COandCP}
$\gamma^\prime_{ij}$'s are nothing but the correlation parameters which were introduced in Ref.~\cite{ref:Stelmchovic}.
\bibitem{ref:gamma'}
We use the symbol $\propto$ between two operators $A$ and $B$ as $A \propto B \Leftrightarrow \exists \ \epsilon > 0 $ such that $A = \epsilon B$. Notice that an operator $\rho_{\text{\tiny{COR}}}$ can be considered as a correlation operator only when the resseting operator $\rho_{\rm{AB}}= \rho_{\rm{A}} \otimes \rho_{\rm{B}} + \rho_{\text{\tiny{COR}}}$ is a density operator with suitable choices of density operators $\rho_{\rm{A}}$ and $\rho_{\rm{B}}$. However, if we choose $\epsilon \le \mathrm{max}_{l,m}[1/(NM||\sigma_l \otimes \tau_m ||)]$ and $\rho_{\text{\tiny{COR}}} = \epsilon \sigma_l \otimes \tau_m$, then the operator $\rho_{\rm{AB}} = \rho_{\rm{A}} \otimes \rho_{\rm{B}} + \rho_{\text{\tiny{COR}}}$ with $\rho_{\rm{A}} = \mathbb{I}_{\rm{A}}/N$ and $\rho_{\rm{B}} = \mathbb{I}_{\rm{B}}/M$ is a density operator of the total system, since (i) $\tr_{\rm{AB}} \rho_{\rm{AB}} = 1$ and (ii) $\forall |\psi\rangle,\ \langle \psi | \rho_{\rm{AB}} | \psi \rangle = ||\psi||^2/NM + \epsilon \langle \psi | \sigma_l \otimes \tau_m | \psi \rangle  \ge ||\psi||^2(1 - \epsilon NM||\sigma_l \otimes \tau_m ||)/NM  \ge 0 $. See Ref.~\cite{ref:Gen1} for the completion of these conditions.
\bibitem{not:VofCNOT}
For the case of the controlled-NOT gate, the initial correlation of Eq.~(\ref{eq:ParticularIC}) does not satisfy these requirement at $m=3$, the fact of which allows non-zero interaction (\ref{eq:VofCNOT}).
\bibitem{not:Inf}
We notice that the proof here is not applied to the case of infinite dimensional composite systems (or those with infinite degrees of freedom). Such a case has been left as an open problem.
\end{thebibliography}
\end{document}